# The geometry of high-dimensional phase diagrams:
## II. The duality between closed and open chemical systems

Jiadong Chen[1], Matthew J. Powell-Palm[2], Wenhao Sun[1]*

[1]Department of Materials Science and Engineering, University of Michigan, Ann Arbor, MI, 48109, USA
[2]J. Mike Walker '66 Department of Mechanical Engineering, Texas A&M University, College Station, TX, 77802, USA
[3]Department of Materials Science & Engineering, Texas A&M University, College Station, TX, 77802, USA

*Correspondence to: whsun@umich.edu

**Significance Statement**

Modern materials are often synthesized or operated in complex chemical environments, where there can be numerous elemental species, competing phases, and reaction pathways. When analyzing reactions using the Gibbs free energy, which has a natural variable of composition, it is often cumbersome to solve for the equilibrium states of a complicated heterogeneous mixture of phases. However, if one is interested only in the stability of a single target material, it may be easier to reframe the boundary conditions around only the target material-of-interest, with boundary conditions open to chemical exchange with an external reservoir. The corresponding phase diagram would thus have a chemical potential axis for the open volatile species, rather than a composition axis. Here we discuss how to derive, compute, and interpret phase diagrams with chemical potential axes, which are dual to the more common composition phase diagram.

**Abstract**

In our ambition to construct high-dimensional phase diagrams featuring any thermodynamic variable on its axes, here we examine the duality between extensive and intensive conjugate variables in equilibrium thermodynamics. This duality manifests in multiple forms, from the distinction between closed and open boundary conditions of a thermodynamic system, to the relationship between the Internal Energy and its Legendre transformations, to the point-line duality in calculating convex hulls versus half-space intersections. As a representative example, here we derive the duality relationships for chemical work with extensive composition variables, $N$, and intensive chemical potentials, $\mu$. In particular, we discuss *mixed* composition-chemical potential diagrams, where some species are volatile while others are not; for oxynitride synthesis, lithium-ion cathode stability, and oxide scale formation on compositionally-complex alloys. We further illustrate how chemical potential diagrams directly connect to materials kinetics, revealing the thermodynamic driving forces for non-equilibrium growth and dissolution processes.

*Duality* is a profound and powerful concept in mathematical physics. As described by Michael Atiyah,[1] duality gives *two different points of view of looking at the same object*. For example, a periodic function can be decomposed into a Fourier series—an infinite sum of sines and cosines; or be represented by its dual the Fourier spectrum—a vector of coefficients for each sine or cosine term. In solid-state physics, Gibbs used duality to invent the concept of the 'reciprocal lattice',[2,3] which is dual to the real-space crystalline lattice—a foundational principle in X-ray diffraction, Bloch wavefunctions, and Ewald summations. A dual representation does not produce any essentially new information, rather, it offers a new perspective to analyze and interpret a physical scenario.

In thermodynamics, there is a duality in how one can ascribe boundary conditions to a thermodynamic system. For a system containing two or more phases coexisting in equilibrium, we can either frame closed boundary conditions around this heterogeneous mixture of equilibrium phases; or if we are interested in the stability of only a single target phase, we could alternatively frame open boundary conditions around only that material, where extensive quantities (heat, volume, mass, *etc.*) are exchangeable with an external reservoir with applied intensive conditions (temperature, pressure, chemical potential, *etc.*). With closed boundary conditions, the relevant phase diagram has the corresponding extensive variable on the axis; whereas for open boundary conditions the relevant phase diagram would have an intensive variable axis. To construct the relevant thermodynamic potential for a phase diagram with natural intensive variables, one uses the Legendre transformation, $\Phi_i = U - \Sigma_i X_i Y_i$, to change the natural variable of a thermodynamic potential from extensive in $U(X_i)$ to a new potential with the conjugate intensive natural variable, $\Phi(Y_i)$.

Previously in Part I of this three-part series on high-dimensional phase diagrams, we showed that Gibbs' physical arguments for heterogeneous equilibrium correspond to the lower convex hulls on the Internal Energy surface, $U(S,X_i)$ of the various possible phases. However, due to Gibbs' stability criterion that $\partial^2 U/\partial X_i^2 > 0$, convex hulls can only calculate heterogeneous equilibrium on thermodynamic axes of extensive variables. In many experimental contexts, the control variables are intensive, like temperature or pressure, which motivates the use of Legendre transformations from $U(S,V,N)$ to the Gibbs free energy $G(T,P,N) = U - (-PV) - TS$. In our ambition to construct generalized high-dimensional phase diagrams with any intensive or extensive thermodynamic variable on the axis, here we discuss the physical interpretation, geometric principles, and computational approaches needed to examine the duality between closed and open thermodynamic systems.

In particular, this work focuses on the duality between closed and open *chemical* systems, where the extensive variable $N$ is mass and the conjugate intensive variable $\mu$ is chemical potential. Although chemical potential diagrams have been previously made in the literature,[4-13] we believe they are underutilized, which we attribute to a lack of literature that describes how to meaningfully interpret chemical potential diagrams. First, we derive how the duality between convex hulls and half-space intersections offer a computational foundation to connect composition phase diagrams to chemical potential diagrams. We then explore how chemical potential diagrams offer a pathway to connect equilibrium thermodynamics to non-equilibrium materials kinetics—as the equations of diffusion, nucleation, growth, and dissolution, all have terms for chemical potential in their constituent equations.

Finally, we will discuss the limitations of phase diagrams with either all composition axes or all chemical potential axes—as there are many physical scenarios where a system is closed in some elements while being open to others. We advocate for the construction of *mixed* composition-chemical potential phase diagrams, with chemical potential axes for the volatile species and composition axes for the closed species. We examine three case studies of mixed intensive-extensive phase diagrams to interpret the synthesis, operational stability, and growth of: 1) the oxynitride TaON, 2) the lithium-ion battery cathode material $LiMn_2O_4$, and 3) oxidation of the medium-entropy alloy $CrCoNiO_x$. To construct these mixed composition-chemical potential diagrams, we combine thermodynamic calculations from both convex hull and half-space intersection algorithms, and discuss the geometry of phase coexistence regions in these mixed diagrams.

More generally, our approach to duality here serves as a general blueprint to Legendre transform the $U(S,X_i)$ convex hull to any high-dimensional phase diagram, either with axes of all intensive variables (such as elastic stress, electric field, magnetic field, surface area to volume ratio, and others), or some mixture of intensive and extensive variables.

### The duality between open and closed thermodynamic systems

It is not meaningful to construct or interpret phase diagrams before establishing the boundary conditions for the thermodynamic system being analyzed. For a material that can undergo chemical reactions, there are two ways to frame boundary conditions, as illustrated in **Figure 1**. When using the Gibbs free energy, which has a natural extensive variable of composition, one frames a *closed* thermodynamic system where the total composition within a reactor is fixed. Inside these closed boundary conditions, an initial non-equilibrium set of reactants will evolve to a final equilibrium phase or a heterogeneous mixture of phases, depending on the ratio of elements in the total reaction vessel. For thermodynamic systems with many phases and chemical species, the resulting network of stoichiometrically-balanced chemical reactions can become very complicated to navigate.[14,15]

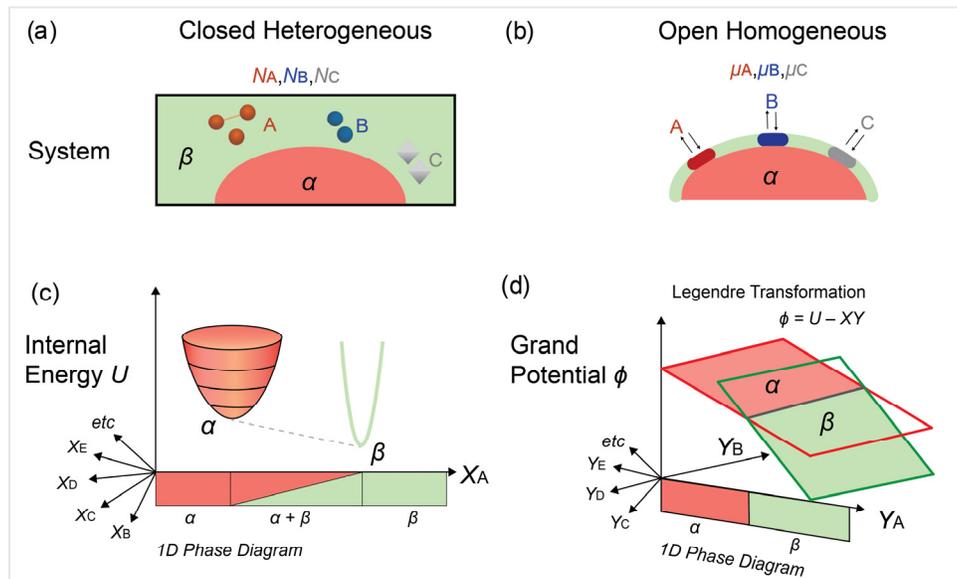

**Figure 1**. **The duality between closed and open thermodynamic systems**; shown in (a, b) with corresponding free energy surfaces (c,d), solved with convex hulls on extensive axes, or half-space intersections on intensive axes.

From a materials engineering perspective, our interest may instead only be with regard to a single target phase—for example, to predict optimal synthesis conditions, or to evaluate operational stability in complex chemical environments. In such cases, we may not need to (or care to) fully characterize all the possible reactions within a closed chemical system. Instead, we could frame the thermodynamic boundaries around only the material-of-interest, which we treat as *open* to an external chemical reservoir that has a propensity to deposit mass onto, or dissolve mass away from, our material-of-interest. The thermodynamic propensity to flux mass onto or away from the target material is given by the chemical potential difference between the reservoir and the material.

A few examples where a thermodynamic system might better be described with open instead of closed boundary conditions include the following: In gas-phase deposition, such as chemical vapor deposition or molecular beam epitaxy, one does not usually control the composition of the volatile species; one controls partial pressure, temperature, and flow-rate, which is better described by chemical potential.[16,17] During precipitation or dissolution from an aqueous electrochemical solution, one usually cares about the material being formed or dissolved, rather than all the various chemical reactions that are possible in $H_2O$.[18,19] In heterogeneous solid-state systems, such as the [cathode | electrolyte | anode] system of an all-solid-state battery, one could examine heterogeneous equilibrium in the convex hull isopleth connecting the cathode and anode;[18] or equivalently one could consider the chemical potential differences at the interfaces between the various electrodes in the battery.[20] Likewise, one can make similar arguments during solid-state synthesis, as one can examine reactions with composition fixed natural variables,[13] or equivalently one could examine the chemical potential differences at the interfaces between reactants and products.[11]

The decision to either model chemical reactions in a closed heterogeneous system, or a collection of subsystems open to each other via chemical exchange, is an arbitrary decision for a human scientist. Nature will evolve the chemical system all the same—but for our conceptual benefit, we should choose our boundary conditions based on whatever considerations are convenient or important to us. To anthropomorphize the target material, it does not 'know' the composition of the reaction vessel—it will simply undergo reactions with the chemical reservoir at its physics interfaces. These reactions proceed until the chemical potentials inside the material are equivalent to the chemical potentials with the reservoir at its interface, such that equilibrium is reached.

## The duality between convex hulls and half-space intersections

Chemical potential diagrams have previously been calculated, most notably by Yokokawa[4], and have been applied to study solid-oxide fuel cells,[5] hydrogen storage materials,[6] surface adsorption,[7,8,9] defects,[9,11] and materials synthesis.[11,13] In our overarching ambition to calculate high-dimensional phase diagrams, it is important to use computational phase diagram approaches that are scalable to many dimensions. Published algorithms for computing chemical potential diagrams, from our perspective, rely on inelegant approaches, often involving for loops or inefficient optimization approaches, and do not scale well to high-component chemical spaces. In addition, published chemical potential diagrams only depict equilibrium situations, while here we derive chemical potential diagrams that can represent non-equilibrium scenarios like crystal growth and dissolution.

Our derivation of the chemical potential diagram below is inspired by the discussions in Callen[21], Zia[22], and Yokokawa[4], but it is derived in a way specifically to leverage the duality between convex hulls and half-space intersections, which are computational optimization algorithms that readily scale to higher dimensions. First, we will connect the Legendre transformation to Point-Line Duality. Then we will use Point-Line Duality to connect convex hulls to half-space intersections. From our perspective, these ideas appear in the literature in scattered form, which our goal here is to unify under one physical, mathematical and computational perspective. For the sake of completeness, we present a full derivation in the **SI1-SI3**. A brief summary of the derivation follows:

For phase diagrams with extensive variables, the equilibrium state is solved by convex hulls, and each state of a phase is given by a vertex on the convex hull. We then leverage a concept from projective geometry named *Point-Line Duality*. Briefly summarized: for a line of the form $y = ax - b$, it is usually customary to treat $x$ and $y$ as the axes, and $a$ and $b$ as parameters for the line. However, since $a$ and $b$ provide all the information needed to define this line, we could equivalently represent this line as a point $(a,b)$ in $a$–$b$ space. Symmetrically, one can swap $a,b$ and $x,y$ to arrive at a similar relationship between lines in $a$–$b$ space and points in $x$–$y$ space. If the line is provided as an inequality, $y \leqslant ax - b$, one can show that the lower convex hull for a collection of points is equivalent to the lower half-space intersection for its dual representation of lines.

In mathematics, the Legendre transformation is a method to relate a convex function to its envelope of tangent lines. Because the tangent line to a $U(X_i)$ surface, $\partial U/\partial X_i$, gives the intensive variable $Y_i$; the Legendre transformation is a natural implementation of Point-Line duality. For a natural intensive variable of chemical potential, the Legendre transformation can be used to construct a new thermodynamic grand potential, $\phi = G - \mu N$. For composition axes, it is customary to transform the number of mols, $N$, to mol fraction, $x$, by the affine relationship $\Sigma_i x_i = 1$, where $x_i = N_i / \Sigma_i N_i$. This changes the intensive variable $\mu$ from the slope $\partial G/\partial N$, to the intercept rule (derivation in **SI3.2**), where $\mu_i = G - (1-x_i) \cdot [dG/dx_i]$. Graphically, $\mu_i$ can be solved by a tangent line of the convex hull, extended to the vertical $G$ axis at the elemental end-point compositions, as later illustrated in **Figure 2a**. Hence, the chemical potential denoted as $\mu$ in this context represents the relative chemical potential to a standard reference state. In some published papers, it is expressed as $\Delta\mu(\mu-\mu^o)$.

The $\phi(\mu)$ space is dual to the $G(x)$ space. Each phase, which was a vertex in $G(x)$ space, becomes a line in the dual $\phi(\mu)$ space. More generally, in higher dimensions, each phase becomes a hyperplane by the equation $\phi = G - \Sigma\mu_i x_i$, which corresponds directly to $y = \Sigma_i a_i x_i - b$. The equilibrium state in $G(x)$ was solved using a bounding lower-convex hull of vertices, which in dual $\phi(\mu)$ space, corresponds to a bounding upper half-space intersection of hyperplanes. A mathematical proof of this duality is provided in **SI2**.

Phase diagrams are constructed by projecting the lowest free-energy phases onto the thermodynamic axes, thus eliminating the energy axis. For example, a *T–P* phase diagram is constructed by projecting the half space intersection of $G = H + PV - TS$ onto the *T* and *P* axes. Likewise, a composition phase diagram is a projection of the $G(x)$ convex hull onto the composition axes; and a chemical potential diagram is a projection of the lower half-space envelope of $\phi$ surfaces onto chemical potential axes (**Figure 2b**). The stability region of a phase on a chemical potential diagram shows chemical potential values where its grand potential is lower than that of any other phase.

**Table 1**. Dual representation of physical aspects in closed system (the convex hull) and open system (chemical potential diagram).

| Physical Aspects | | Dual Representation | |
|---|---|---|---|
| **System** | **Boundary Condition** | **Closed to mass transfer** | **Open to mass transfer** |
| Thermodynamic Aspects | Thermodynamic Potential | Gibbs potential | Grand potential |
| | Euler form | $G = U + pV - TS$ | $\Phi = G - \mu N$ |
| | Differential Form | $dG = -SdT + vdP + \mu dN$ | $d\Phi = -SdT + vdP - Nd\mu$ |
| | Natural Variables | $T, P, N$ | $T, P, \mu$ |
| | Heterogeneous Equilibrium | Coexistence region | Phase boundary |
| | Metastability | Energy above the hull | Growth |
| | Instability | Decomposition energy | Dissolution |
| Geometric aspects | Phase | Point | Hyperplane |
| | Equilibrium state | Convex Hull | Half-space intersection |

In this work, we will present chemical potential diagrams for a variety of systems. All the thermochemical data for these diagrams are from the Materials Project database[23], which is a database of high-throughput DFT-calculated enthalpies of ordered crystalline phases. As is common in the computational materials science community, we assume that vibrational entropy is negligible in solids, so that we can approximate $G_{\text{solid}} = E_{\text{Solid,DFT}}$[24]. The Materials Project only contains ordered crystalline phases, so we do not consider the solid-solution phases, although they are certainly relevant in real materials. Finally, DFT has known errors in formation energy[25,26], but the Materials Project has implemented a series of energetic corrections[27,28,29] which we adopt here without further scrutiny. For a thorough analysis of the actual chemical systems presented here, it would be appropriate to recalculate the thermochemical data without these assumptions—however this work primarily emphasizes the formalism, geometry and interpretation of chemical potential diagrams, so we use the unaltered Materials Project data for our visualizations.

## Equilibrium and non-equilibrium regions on a chemical potential diagram

When analyzing the stability of a material under open boundary conditions, one should distinguish in the mind between the *internal* intensive variable of a substance, $Y_i$, versus the *external* intensive variable of the reservoir $Y_{external}$. If the $Y_{internal} \neq Y_{external}$, then the conjugate extensive quantity $X$ will flow through the boundary until $Y_{internal} = Y_{external}$, after which entropy will be maximized and equilibrium is reached. For example, if 50°C water is exposed to an external temperature reservoir of 10°C, heat will flow out of water into the reservoir, and the entropy of water will be reduced correspondingly. Water has a continuous span of entropies in this temperature range, so it can change its *internal* extensive entropic state to equilibrate with an *external* intensive temperature reservoir.

Likewise, for chemical work, to equilibrate the internal chemical potential of a material with the external chemical potential of the reservoir, mass can be transferred across the boundaries. At **equilibrium**, the externally applied chemical potential reservoir will be exactly equal to the internal chemical potential of a material; $\mu_{material} = \mu_{reservoir}$, such that there is no driving force to transfer mass to or from the reservoir. For a ***non-equilibrium*** situation, if external chemical potentials are different than the internal chemical potentials of a material, mass will have a propensity to flux from high $\mu$ to low $\mu$; where a material will grow if $\mu_{reservoir} > \mu_{material}$, or it will dissolve or corrode if $\mu_{reservoir} < \mu_{material}$. Because growth and dissolution are fundamental aspects of materials kinetics, chemical potential diagrams offer a direct link between non-equilibrium thermodynamics and kinetics of transport.

The internal chemical potential of the material derives from the energies of its quantum-chemical and electrostatic bonds—which determines its scalar formation energy. **Figure 2a** visualizes the formation energies of phases from the Mn-O system with its interpretation for equilibrium and non-equilibrium scenarios from the perspective of a convex hull. The corresponding chemical potentials can be interpreted from the intercept rule—where $\mu_{Mn}$ or $\mu_O$ are the intercept of the tangent lines of the convex hull with the vertical energy axis at the elemental end-point compositions.

The **equilibrium** chemical potential window of the single phase $Mn_3O_4$ is bound between the cotangent lines of $Mn_2O_3/Mn_3O_4$ and $Mn_3O_4/MnO$, where these cotangent lines indicate the chemical potentials where $Mn_3O_4$ can coexist in equilibrium with $Mn_2O_3$ or $Mn_3O_4$. **Figure 2b** shows the corresponding chemical potential windows of each $Mn_xO_y$ phase, indicated by the vertical line segments at a given composition. Because phases on the convex hull are points, the Legendre transformation of these phases form the grand potential surfaces, whose half-space intersection is shown **Figure 2d**. The condition where the *externally* applied chemical potentials are equal to the *internal* chemical potential of a material can be written as:

$$\phi = [G]_{internal} - [\Sigma_i \mu_i x_i]_{external} = 0.$$

Therefore, the conditions of equilibrium on a chemical potential diagram correspond to a *slice* of grand potential surfaces where $\bar{\phi} = 0$, accentuated by the darker lines on **Figure 2d**. We call this the *equilibrium envelope* of the chemical potential diagram. All regions in chemical potential diagrams display the external chemical potential applied by the reservoir, but darker lines additionally represent the situation where *internal* chemical potential of each material on the convex hull is equal to the external chemical potential from the reservoir. The lines formed by the equilibrium envelope ($\phi = 0$) are consistent with the vertical segments formed by intercept rule as illustrated in **Figure 2b**.

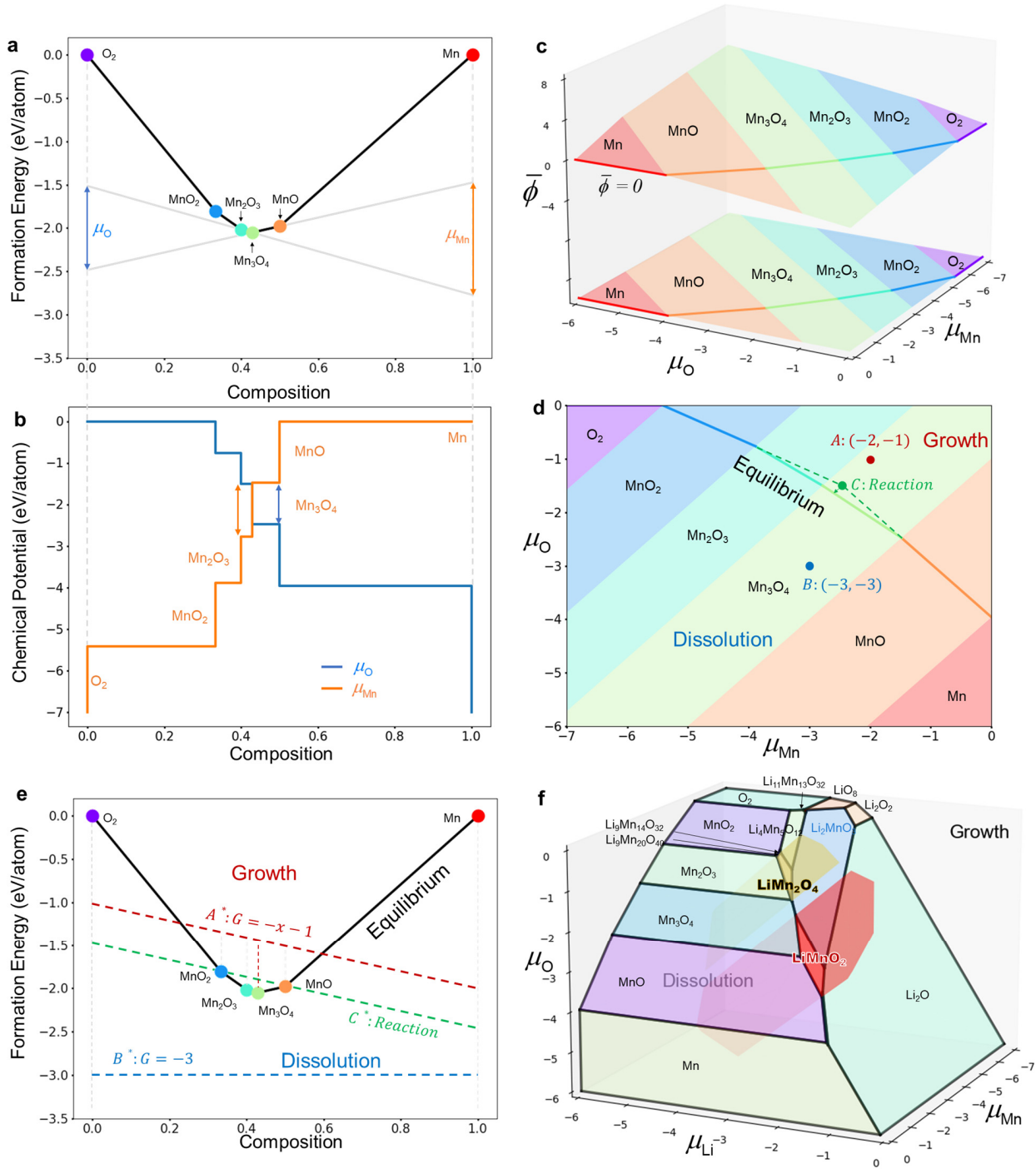

**Figure 2**. Duality between convex hulls and chemical potential diagrams in the binary Mn-O system. (**a**) Tangent lines to the convex hull, and their intercepts with the energy axes, show the elemental chemical potential window for $Mn_3O_4$. (b) Chemical potential windows for $\mu_{Mn}$ and $\mu_O$ various $MnO_x$ phases indicated by vertical segments. (c) Grand potential hyperplanes for the Mn-O chemical potential diagram. The equilibrium envelope is emphasized by a dark line at $\phi = 0$ (d) Dissolution and growth regimes on the chemical potential diagram and (e) their dual relationship with the convex hull. (f) Ternary chemical potential diagram, with the growth and dissolution regimes for $LiMn_2O_4$ and $LiMnO_2$ illustrated as extending in and out of the stability regions on the equilibrium envelope.

The equilibrium envelope further separates a chemical potential diagram into non-equilibrium regions of growth and dissolution. For example suppose $Mn_3O_4$ is placed in contact with an external chemical reservoir where the boundary conditions are $(\mu_{Mn}, \mu_O)_{external} = (-2, -1)$, indicated by the red label A in **Figure 2d**. These chemical potentials are higher than the internal $\mu_{Mn}$ and $\mu_O$ in $Mn_3O_4$, so Mn and O will flux from the external chemical reservoir onto $Mn_3O_4$, leading to crystal growth. On the convex hull in **Figure 2e**, the dual representation to this point on the chemical potential diagram corresponds to the line $A^*$ on the convex hull. Likewise, if $Mn_3O_4$ is exposed to low $(\mu_{Mn}, \mu_O)_{external} = (-3, -3)$, shown as point B in **Figure 2d** and the line $B^*$ in **Figure 2e**, Mn and O will flux out of $Mn_3O_4$ into the reservoir, leading to dissolution of $Mn_3O_4$. The precise chemical or structural nature of the external chemical reservoirs are irrelevant, only their $\mu_{Mn}$ and $\mu_O$ chemical potentials matters.

The chemical potential diagram can also show conditions for solid-solid phase transformation from a metastable solid to an equilibrium phase. If a different $MnO_x$ phase, for example $Mn_2O_3$, were exposed to an external chemical potential such as point A or B, which is in the non-equilibrium region corresponding to $Mn_3O_4$, there would first be a thermodynamic driving force for phase transformation from $Mn_2O_3$ to $Mn_3O_4$, followed by subsequent growth or dissolution of $Mn_3O_4$. In the convex hull of **Figure 2e**, the phase transformation that originates from $Mn_3O_4$ has the largest energy drop from line $A^*$ compared to all other $MnO_x$ phases. Nucleation and diffusion kinetics aside, the bulk driving force preferences the transformation and further growth of $Mn_3O_4$. Additionally, because $Mn_3O_4$ has the shallowest energy drop to $B^*$, any other $MnO_x$ composition exposed to this external chemical reservoir can first reduce its free energy by transforming to $Mn_3O_4$, and then dissolving mass out to the reservoir.

Additionally, chemical potential diagrams can be employed to illustrate solid-state reactions. In **Figure 2e**, the green dashed line, denoted as $C^*$, represents the reaction between $MnO_2$ and MnO. By Point-Line duality, this corresponds to point C in the chemical potential diagram (**Figure 2d**), which is the intersect between the extension of the equilibrium envelope lines for $MnO_2$ and MnO. The reaction driving force is the distance from point C to the $Mn_3O_4$ equilibrium envelope along the $\mu_O : \mu_{Mn} = 1$ direction. This reaction energy on the chemical potential diagram is equal to the distance between $C^*$ and $Mn_3O_4$ on the convex hull.

Our geometric interpretation of the binary convex hull and chemical potential diagram can be readily extended to higher component systems. **Figure 2f** illustrates a ternary $\mu_{Li}$-$\mu_{Mn}$-$\mu_O$ chemical potential diagram. The chemical potential diagram exists in 3 dimensions, since one can vary all three chemical potentials independently for non-equilibrium scenarios. However, the equilibrium envelope is still a 2-dimensional manifold, due to the special constraint that $\bar{\phi} = [G]_{internal} - [\Sigma_i x_i \mu_i]_{external} = 0$. Nonetheless, one can see that the non-equilibrium regions similarly extend into and out of the equilibrium manifold, for example as highlighted for $LiMn_2O_4$ (yellow) and $LiMnO_2$ (red). All other arguments of growth and dissolution can be applied to this ternary chemical potential diagram.

## Mixed Composition-Chemical Potential Diagrams

There are many physical situations where a chemical system is open to some elements, but are closed in others. For example, in the stability of metal oxynitrides, oxygen and nitrogen can be volatile, whereas the metal(s) usually are not. In such cases, pure chemical potential diagrams or pure compositional phase diagrams may not be the most useful. Here we advocate for the construction of mixed composition and chemical potential phase diagrams, interpreting three representative case studies for oxynitride stability, lithium-ion cathode stability, and oxidation of multicomponent alloys.

Although mixed composition-chemical potential diagrams can be interpreted from the geometry of the intercept rule, they are not straightforward to calculate, as they require using convex hulls in the composition axes, and then half-space intersections for the chemical potential axes. We present a method where equilibrium is calculated both with convex hulls and half-space intersections, then the coordinates of each phase are mixed-and-matched depending on if the desired axis is composition or chemical potential. Details of this computational implementation are discussed in **SI4**.

### Metal oxynitrides

Oxynitrides are a class of mixed-anion materials with applications for semiconductors and optoelectronics,[30] water-reduction photocatalysts,[31] electrocatalytic nitrogen reduction,[32] hard coating,[33] energy storage,[34,35] *etc*. Introducing additional anions with different sizes, electronegativities, and charges can effectively modulate the physical properties of oxide-based compounds[36,37]. However, oxynitrides are difficult to synthesize, and if synthesized, do not always retain operational stability (for example during catalysis in harsh electrochemical environments). Here, we examine the boundary conditions and relevant phase diagram in evaluating the stability of tantalum oxynitride, TaON.

It is not straightforward to experimentally control the oxygen and nitrogen *composition* in a reaction vessel, as oxygen and nitrogen are gases at standard state and at elevated temperatures. For this reason, it is not very convenient to examine oxynitride stability on phase diagrams with oxygen and nitrogen *composition* axes. If we are only concerned about the stability of the oxynitride, we can instead frame our thermodynamic system around just the oxynitride itself, with boundary conditions open to oxygen and nitrogen transfer, while closed in the non-volatile metal species. The corresponding phase diagram should therefore be a mixed $x_{metal}$-$\mu_O$-$\mu_N$ diagram.

In **Figure 3a**, we illustrate the geometric connection between an all-extensive $x_{metal}$-$x_O$-$x_N$ convex hull with its mixed $x_{metal}$-$\mu_O$-$\mu_N$ phase diagram. For a target TaON phase, the blue triangle indicates the tangent plane to the TaON vertex. The intercept of this tangent plane with the energy axes at the pure elemental compositions corresponds to the elemental chemical potentials. Tilting this tangent plane about the TaON vertex maps out $\mu_O$ and $\mu_N$ chemical potentials where TaON is a stable equilibrium phase. This tangent plane tilting process is similar to retrieving the temperature and pressure of a phase on the Maxwell $U$-$S$-$V$ surface, except that on the Maxwell surface the slope of the tangent plane $\partial U/\partial X$ directly gives the intensive variable $Y$, whereas in affine composition axes (where $x_1 = 1 - x_2 - x_3$), the conjugate intensive chemical potential variable is given by the intercept rule.

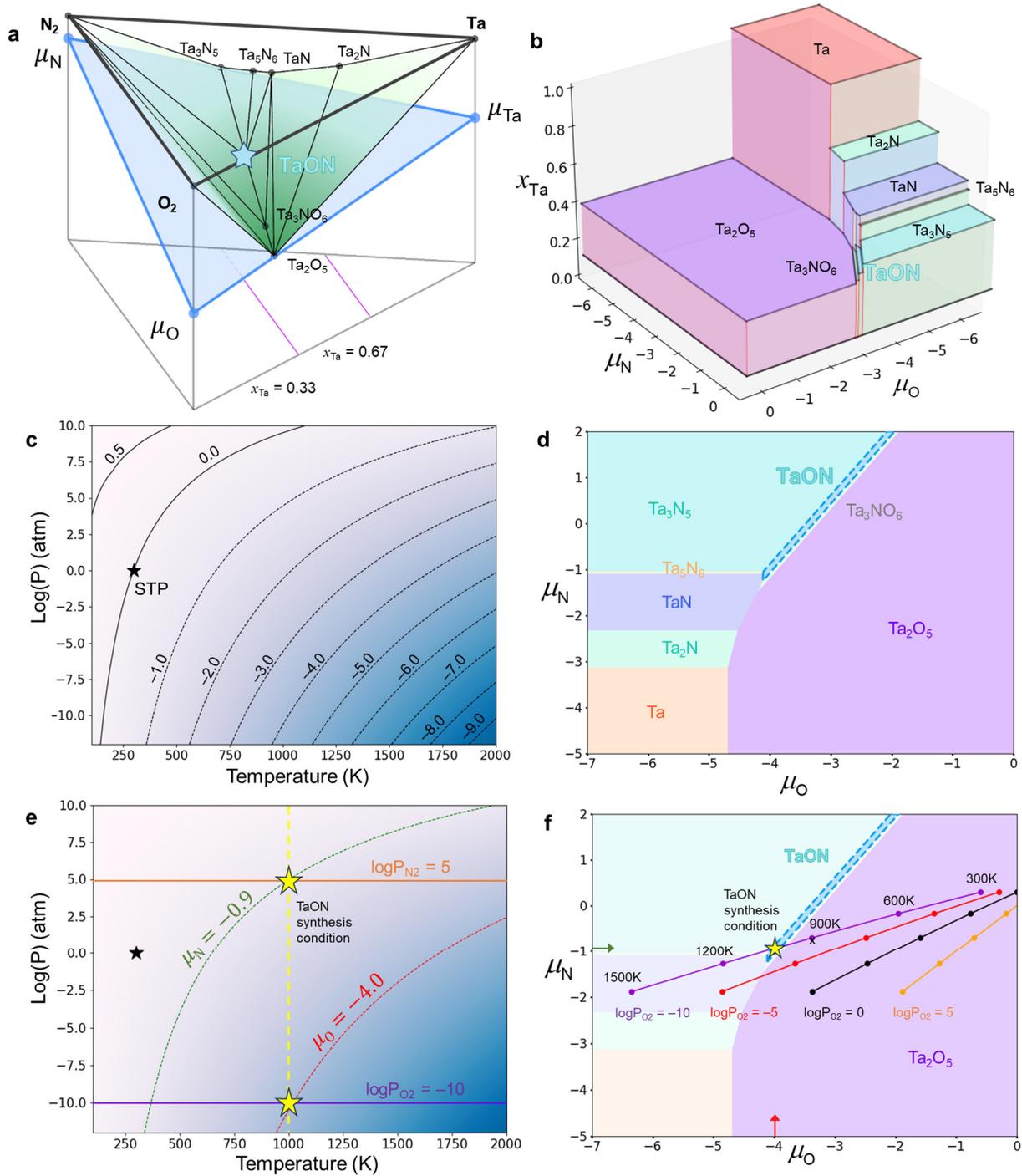

**Figure 3**. **a)** Ternary $x_{Ta}$-$x_N$-$x_O$ convex hull. Blue triangle represents tangent plane to the TaON vertex, whose intercepts with the energy axis provides the corresponding elemental chemical potentials. **b)** Mixed $x_{Ta}$-$\mu_O$-$\mu_N$ phase diagram. **c)** The chemical potential of a diatomic gas like $O_2$ or $N_2$, as a function of temperature and partial pressure. Iso-µ lines are marked from -9 to 0.5 eV/atom. **d)** $\mu_O$-$\mu_N$ projection of the mixed $x_{Ta}$-$\mu_O$-$\mu_N$ diagram. **e)** Gas conditions for $N_2$ and $O_2$ where TaON is stable, marked with yellow star. **f)** Lines on the $\mu_O$-$\mu_N$ projected diagram corresponding to gases at various partial pressures, where we fixed $logP_{N_2} = 5$, and then show isolines corresponding to $\mu_O$ and $\mu_N$ at various $logP_{O_2}$ and temperature.

The $x_{Ta}$-$\mu_O$-$\mu_N$ phase diagram is shown in **Figure 3b**, where single-phase regions correspond to horizontal polygons with black borders parallel to the $\mu_O$ and $\mu_N$ axis. 2-phase coexistence regions are formed by the vertical rectangles connecting two single-phase polygons parallel to the $x_{Ta}$ direction, The 3-phase coexistence regions are given by the vertical red lines that connect two 2-phase coexistence regions. The $x_{Ta}$ axis shows how changing $\mu_O$ and $\mu_N$ can control the Ta molar fraction. However, this 3-dimensional diagram can also be projected onto just the $\mu_O$-$\mu_N$ axes, as in **Figure 3d**.

The chemical potential of a gaseous phase is given by $\mu_{gas} = \mu_0 + RT\ln[P_{gas}] - TS_{gas}$, where $\mu_0$ is the standard state chemical potential, $P_{gas}$ is the partial pressure, and $S_{gas}$ is its entropy. $N_2(g)$ and $O_2(g)$ are the equilibrium elemental phases at standard state, so for both oxygen and nitrogen, $\mu_0 = 0$ at 298K and $P = 1$ atm. The dependence of $\mu_{gas}$ on temperature and partial pressure is schematized in **Figure 3c**, which provides an experimental reference guide that can be used together with the $x_{metal}$-$\mu_O$-$\mu_N$. The combination of **Figures 3b**, **c**, **d** provides theoretical utility similar to the Ellingham diagram, however the Ellingham diagram cannot examine materials stability with two independent volatile gaseous species, whereas the chemical potential diagram construction can.

Although TaON is on the Ta-O-N convex hull, and is therefore a thermodynamically stable phase, its stability window is very narrow in the $\mu_O$-$\mu_N$ diagram, meaning that the conditions to stabilize TaON may need to be very precise. In particular, $\mu_N$ should be much greater than $\mu_O$ for TaON to be stable. In **Figure 3f**, we place a yellow star in the TaON stability window at $(\mu_O, \mu_N) = (-4$ eV, $-0.9$ eV$)$. By referencing the diagram in **Figure 3e**, we can determine the $O_2$ and $N_2$ gas conditions that correspond to this TaON stability point.

For an oxynitride exposed $O_2(g)$ and $N_2(g)$, the temperatures of the two gases will be the same—however, their relative partial pressures can be varied independently. On **Figure 3e**, we should search for a temperature (a vertical line) that intersects iso-$\mu$ lines of $-0.9$ eV for $N_2$, and $-4$ eV for $O_2$. One such condition is at 1000K, with $\log P_{N2} = 5$ and $\log P_{O2} = -10$. In **Figure 3f**, we use a series of dotted lines to represent different temperatures and different $\log P_{O2}$, with each line having a fixed $\log P_{N2} = 5$. For most conditions, these lines fall in $Ta_2O_5$ region, showing that TaON is unstable with respect to $Ta_2O_5$ under most conditions in air. However, for the line $\log P_{O2} = -10$, we can intersect the TaON region at 1000K.

Although we conducted our stability analyses with respect to $O_2$ and $N_2$ gas, we can use other nitrogen or oxygen precursors to shift the $\mu_N$ and $\mu_O$ chemical potentials. For example, to overcome the triple bond in the $N_2$ molecule, nitrides are usually much more readily synthesized with activated nitrogen precursors, such as ammonia where the half reaction $\mu_N = \mu_{NH3} - 3/2\ \mu_{H2}$ yields $\mu_N = 0.4$ eV at standard state; and plasma-cracked atomic nitrogen has been benchmarked to $\mu_N = 1$ eV/N.[38,39] These chemical potentials are equivalent to $N_2(g)$ partial pressures of $10^{16}$ atm and $10^{40}$ atm, respectively. A low oxygen chemical potential can also be obtained by reducing agents, for example, reduction with carbon monoxide yields an equilibrium chemical potential of $\mu_O = \mu_{CO2} - \mu_{CO} = -2.6$ eV/atom at STP, equivalent to an $O_2(g)$ partial pressure of $10^{-104}$ atm (assuming the reaction is not kinetically-limited). Similar analyses can be done to obtain the effective chemical potential of chemical species in various other states, including solvated aqueous ions, or atoms in other solids.

The $x_{Ta}$-$\mu_O$-$\mu_N$ chemical potential diagram can also yield other insights that cannot be readily obtained from a compositional phase diagram. **Figure 3d** show that to reduce $Ta_2O_5$ to metallic Ta, $\mu_O$ needs to be below -4.6 eV, which are also conditions generally needed to synthesize pure tantalum nitrides. The phase boundary between the various tantalum nitrides $TaN_x$ and the $Ta_2O_5$ also indicates conditions for the stability of the pure tantalum nitrides in air.

**Stability of the Li-ion cathode material LiMn$_2$O$_4$**

$LiMn_2O_4$ is a candidate cathode material for rechargeable Li-ion batteries, in particular because Mn is not a critical element like cobalt-based battery electrodes.[40,41] $LiMn_2O_4$ has a spinel crystal structure with diffusion channels that enable fast diffusion of $Li^+$, even at relatively low concentration of $Li^+$.[42,43] However, during the synthesis and electrochemical operation of $LiMn_2O_4$, many competing ternary $Li_xMn_yO_z$ phases can form, such as $Li_4Mn_5O_{12}$ and $LiMnO_2$, as well as the solid-solution phases that can form between these ternary phases and $MnO_2$, $Mn_3O_4$. (**Figure 4a,b**). The complexity of the available phases and structural transformations, especially between layered rocksalt structures and the spinel structures, can result in undesired phases in the form of impurities during synthesis, as well as irreversible decomposition pathways during electrochemical cycling and operation.

Under various synthesis or operation contexts, all 3 elements Li, Mn and O can be volatile in $LiMn_2O_4$. The oxygen chemical potential can be controlled by an oxidizing or reducing environment during synthesis, and likewise thermal decomposition by metal reduction and oxygen evolution also depends on $\mu_O$[44] During battery charging and discharging, Li is cycled in and out of $LiMn_2O_4$ through the electrolyte,[20] where $\mu_{Li} = \mu_{Li,\text{metal}}^o - e\varphi$,[18] where $\varphi$ is the electric potential, and $\mu_{Li,\text{metal}}^o = 0$ because $\mu$ is referenced to the chemical potential to elemental Li. Because the electrolyte is adjacent to $LiMn_2O_4$, the electrolyte can exchange Li, Mn or O with $LiMn_2O_4$. In particular, one major issue hampering the widespread adoption of manganese-based cathodes is dissolution of the redox-active Mn ion in organic electrolytes, where Mn diffuses through the electrolyte to form an undesirable solid-electrolyte interface (SEI) at the anode, which erodes overall battery capacity.[45]

Although all three elements can be exchanged through an open boundary condition, it can be confusing to analyze $LiMn_2O_4$ stability on a full $\mu_{Li}$-$\mu_{Mn}$-$\mu_O$ chemical potential diagram, since it becomes difficult to isolate the work of the reservoir on the individual volatile species. It may be better to close the system to two components, and examine the role of the reservoir chemical potential on just the third component. From **Figure 4a** through **Figure 4d**, we illustrate how to interpret an $x_{Li}$-$x_{Mn}$-$\mu_O$ diagram from the Li-Mn-O ternary convex hull. Each slice in **Figure 4a, b** is an isopleth between oxygen and a fixed $Li_xMn_{1-x}$ ratio. The Li:Mn isopleth with a 1:2 ratio (purple), corresponding to $LiMn_2O_4$, intersects both pure phases as well as 2-phase tie lines. By viewing the convex hull along this isoplethal slice in **Figure 4c**, we can use the intercept of tangent lines against the $\mu_O$ axis to illustrate the different phase transition and phase-coexistence regions. By repeating this process for all $Li_xMn_{1-x}$ ratios, we can construct the full $x_{Li}$-$x_{Mn}$-$\mu_O$ diagram in **Figure 4d**. The right-side axis of **Figure 4c** shares the same color correspondence with the stability and coexistence regions in **Figure 4d**.

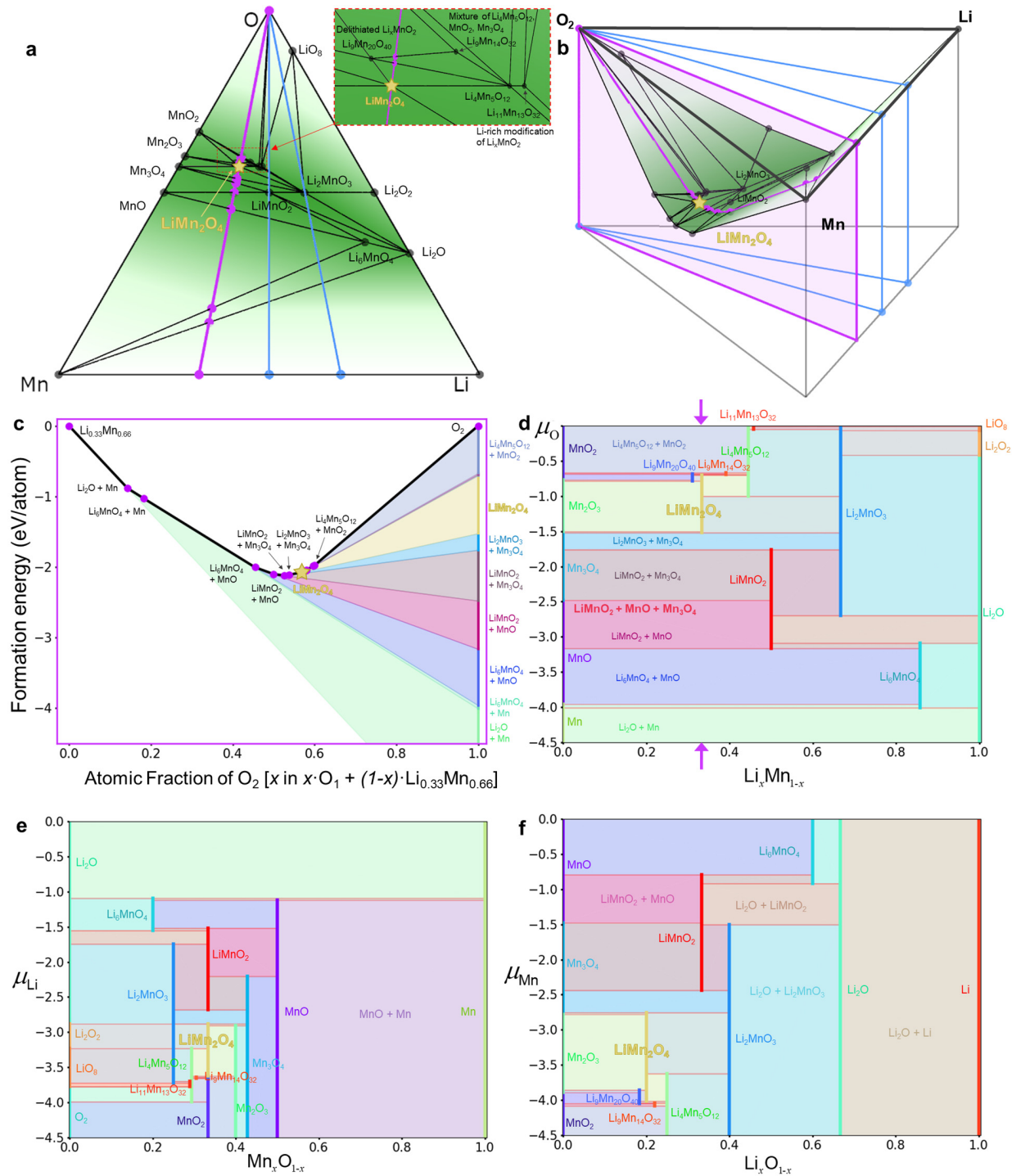

**Figure 4.** **a)** Ternary $x_{Li}$-$x_{Mn}$-$x_O$ convex hull. Inset shows DFT calculated-phases with ordered compositions, where $Li_9Mn_{20}O_{40}$ phase represents delithiated $Li_xMnO_2$, the lithium-rich modification of $LiMn_2O_4$ is represented by $Li_{11}Mn_{13}O_{32}$, and the $Li_9Mn_{14}O_{32}$ phase represents a tie-line between $MnO_2$-$Li_4Mn_5O_{12}$. Isopleths between $O_2$ to $Li_xMn_{1-x}$ are shown, with a purple highlight for a ratio of Li:Mn = 1:2. **b)** Ternary convex hull with energy axis, with isoplethal slices shown. **c)** Intercept rule construction of stability regions and phase coexistence along the $\mu_O$ axis. **d)** mixed $x_{Li}$-$x_{Mn}$-$\mu_O$ phase diagram, **e)** mixed $x_{Mn}$-$x_O$-$\mu_{Li}$ phase diagram, **f)** mixed $x_{Li}$-$x_O$-$\mu_{Mn}$ phase diagram. For the phase coexistence in $x_1$-$x_2$-$\mu_3$ diagram, single phases are vertical lines, 2-phase coexistence regions are rectangles that connects two single phases, and 3-phase coexistence regions are horizontal red lines that connects the ends of three single phases.

From the $x_{Li}$-$x_{Mn}$-$\mu_O$ in **Figure 4d**, we can examine reactions involving $LiMn_2O_4$ with $O_2$ gas, for example during synthesis or thermal decomposition. The $\mu_O$ for $O_2$ gas can be referenced to different temperatures and partial pressures using the earlier diagram from **Figure 3c**. The $\mu_O$ stability window for $LiMn_2O_4$ is between [-1.5, -0.6] eV/atom; corresponding to temperature range around 600 – 900K, for $P_{O2}$ ranges from [0.21, 1] atm, corresponding to ambient atmosphere. This stability condition is in line with the reported solid-state synthesis temperatures of $LiMn_2O_4$, which range from 700 – 1000K[46,47], as well as the thermal decomposition temperature of $LiMn_2O_4$ at 1100K.[48] Additionally, compared to layered structure electrodes, such as $LiMnO_2$, higher $\mu_O$ is beneficial for the stability of $LiMn_2O_4$, which matches the experimental fact that the $Mn^{3+}/Mn^{4+}$ redox in $LiMn_2O_4$ requires a larger amount of oxygen redox to achieve high capacity, compared to $Mn^{2+}/Mn^{4+}$ in $LiMnO_2$[49].

To analyze lithiation process of $LiMn_2O_4$ for a given Mn:O ratio, we can utilize the $x_O$-$x_{Mn}$-$\mu_{Li}$ axis. As shown in **Figure 4e**, when we raise the voltage (thereby decreasing $\mu_{Li}$) to charge $LiMn_2O_4$, it undergoes oxidation and transforms into $MnO_2$. We note that on the equilibrium phase diagram, the $MnO_2$ phase corresponds to the ground-state $\beta$ (pyrolusite) phase, whereas for the real $LiMn_2O_4$ system, topotactic delithiation results in metastable $\lambda$-$MnO_2$, which maintains the spinel framework. On the other hand, reducing the electrostatic potential (increasing $\mu_{Li}$) can result in phase separation to $Li_2MnO_3 + Mn_3O_4$. The $\mu_{Li}$ window between -3.7 and -2.8 eV/atom corresponds to the phase transitions between $MnO_2$ and $LiMn_2O_4$, as well as the transition from $LiMn_2O_4$ to $LiMnO_2$.

For electrolyte stability, Mn dissolution from $LiMn_2O_4$ will occur if the $\mu_{Mn}$ is lower in the electrolyte than the lower-limit $\mu_{Mn}$ stability window in $LiMn_2O_4$. For example, dissolution of Mn occurs if the applied $\mu_{Mn}$ in the electrolyte is below -4.1 eV, which as shown in **Figure 4f**, can induce multiple phase transformations to $Li_4Mn_5O_{12}$, $Li_9Mn_{20}O_{40}$, $Li_9Mn_{14}O_{32}$. To design an organic electrolyte that is resistant to Mn-dissolution, one needs to identify an organic electrolyte where the Mn-ion solvation energy overlaps the stability window of $LiMn_2O_4$ in the $\mu_{Mn}$ axis. To perform this analysis, one can construct the corresponding chemical potential diagram of the electrolyte from a convex hull using the same tangent line principles discussed here.

**Oxidation of compositionally-complex alloys**

Compositionally-complex alloys (CCAs) have near equimolar concentrations of multiple metal species, and in special cases form single-phase solid-solution high-entropy alloys (HEAs) and medium-entropy alloys (MEAs), which may have valuable properties for high-temperature materials for spacecraft and satellites,[50] corrosion-resistance for seawater treatment equipment,[51,52] superior electron transport for electronic device, *etc*.[53,54,55] However, discontinuous oxide granules or oxide layers can form when these alloys are exposed to $O_2$ atmospheres and high temperature. Although experimental measurements of the surface oxide phases formed in HEAs and MEAs are becoming more numerous,[56,57] thermodynamic modeling remains sparse.[58] This may be due to the complexity of the possible binary, ternary, and quaternary oxides that compete to form during HEA/MEA oxidation.

To analyze the oxidation behaviors of multi-component alloys, we take $CrCoNiO_x$ as a representative example. The chemical potential of oxygen in an HEAs depends on many factors, such as penetration depth of oxygen as it diffuses in, as well as the $\mu_O$ applied by the temperature and partial pressure of $O_2$ gas at the surface. The appropriate boundary conditions for this system are closed in the metal species, but with open exchange of volatile oxygen species. The relevant phase diagram is therefore closed with metal composition axes $x_{Cr}$-$x_{Co}$-$x_{Ni}$, and open with a $\mu_O$ axis. For the sake of visualization, we examine a here 3-metal MEA, however, the underlying geometric arguments and analyses are readily extendable to higher-component alloys.

Typically, phase diagrams for 4-component systems are viewed in barycentric coordinates using a 3D Gibbs tetrahedron to represent the quaternary convex hull, with 4 composition variables but no energy axis (**Figure 5a**). The energy axis can be recovered by constructing pseudo-ternary convex hulls by taking compounds as terminal points of the convex hull, and plotting the formation energy of phases relative to the terminal compounds, as illustrated in **Figure 5b**.

Each phase in the quaternary convex hull is a vertex, which we assign a color corresponding to the metal composition. We assign red, green, and blue to Co, Cr, and Ni, respectively. The color of binary through quaternary phases are then determined by their barycentric Co:Cr:Ni molar ratio. We use color saturation to correspond to the lowest critical oxygen chemical potential in which the phase is thermodynamically preferred to form, where white indicates pure $O_2$ gas.

**Figure 5c**, **5d** shows two $x_{Cr}$-$x_{Co}$-$x_{Ni}$-$\mu_O$ phase diagrams at high and low $\mu_O$ ranges—split up to more clearly visualize the phase coexistence regions. By comparing the critical oxygen chemical potentials, we can extract the tendency of various metal constituents to oxidize; for example, Cr will oxidize at $\mu_O$ = -4 eV to form a protective $Cr_2O_3$ scale, which is before Ni and Co which both oxidize around $\mu_O$ = -2.5 eV. Experimentally, the oxidation of equimolar CrCoNi is shown to form only a $Cr_2O_3$ layer[53,54,56], as anticipated by these diagrams. Additionally, a mixed spinel $(Co,Ni)Cr_2O_4$ is experimentally observed[53,54,56,59], which in our phase diagram on **Figure 5c** may correspond to a solid-solution that would form along the tie line between $CoNi_2O_4$ and $CrNiO_4$.

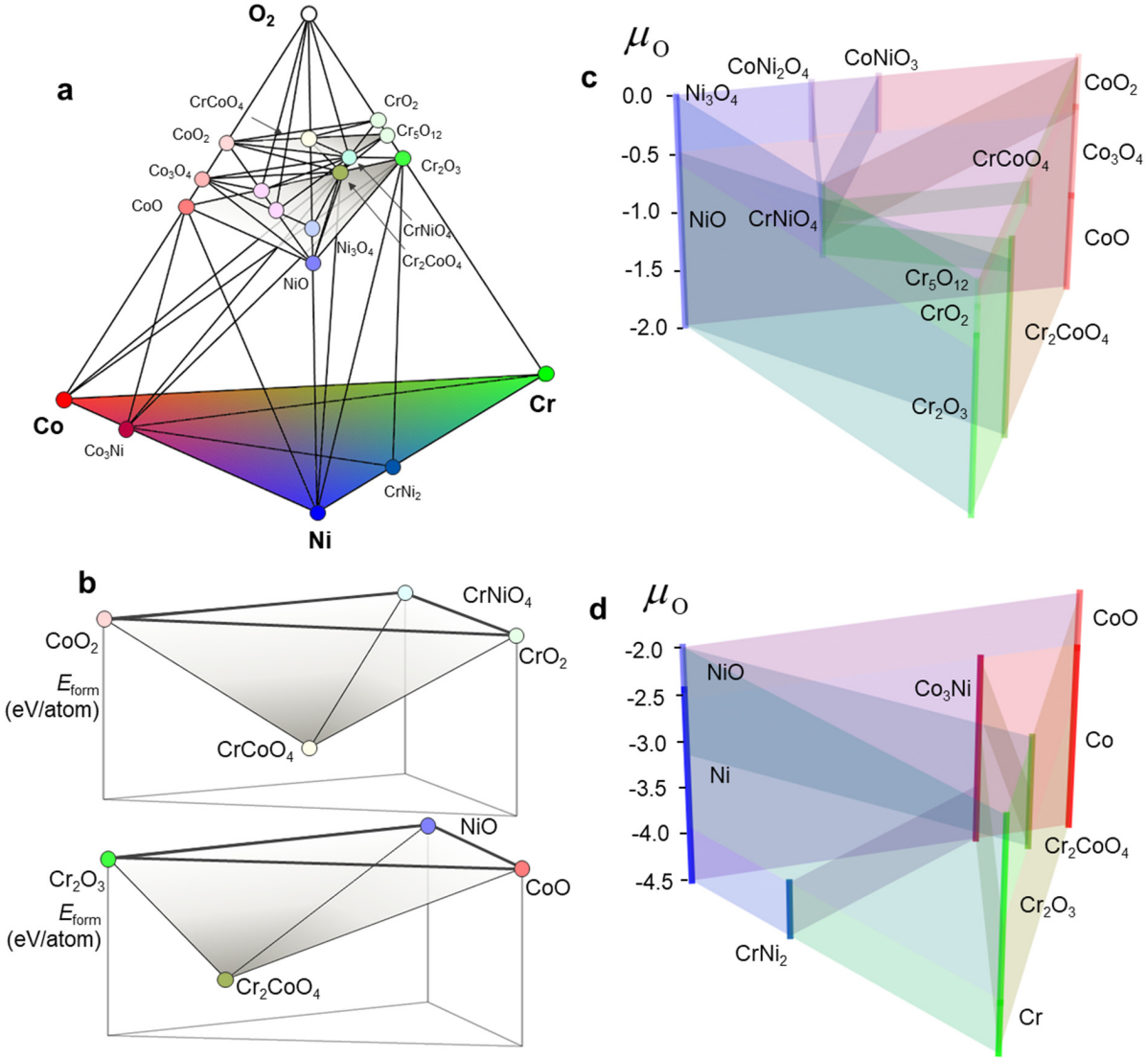

**Figure 5**. **a**) Quaternary $x_{Co}$-$x_{Cr}$-$x_{Ni}$-$x_O$ convex hull. Each single phase is assigned a color based on metal composition of Cr:Co:Ni ratio, and a transparency based on the lowest critical $\mu_O$ for a given phase. Two triangular ternary isopleths are shown in gray colorscale, connecting $CoO_2$-$CrNiO_4$-$CrO_2$, and $Cr_2O_3$-NiO-CoO. **b**) Pseudo-ternary convex hulls with a recovered formation energy axis, with energies referenced to the terminal compound phases. **c**). Mixed $x_{Cr}$-$x_{Co}$-$x_{Ni}$-$\mu_O$ phase diagram, where **c**) $\mu_O$ in [-2.0, 0.0] eV/atom, and **d**) $\mu_O$ in [-4.5, -2.0] eV/atom.

In **Figure 5c**, **5d**, each single phase is a vertical line, a 2-phase coexistence region is vertical rectangle plane bounded by two single phase lines, a 3-phase coexistence region is a triangular prism formed by three single phase lines, and finally a 4-phase coexistence region is a horizontal triangle connected by two 3-phase coexistence triangular prisms. Although 2-phase coexistence regions and 4-phase coexistence regions are both 2-D manifolds, they have different physical interpretations because there is no degree of freedom for changing $\mu_O$ in 4-phase coexistence, leading to a horizontal 2D region, whereas we can change $\mu_O$ in 2-phase coexistence, leading to a vertical 2D region.

**Dimensionality of Coexistence Regions in Mixed Diagrams**

As discussed in Part I of this 3-part series, the essential geometric object corresponding to phase coexistence is the simplicial polytope, which is an $N$-dimensional analogue of a triangle. The counting relations between the vertices, edges, and facets of a simplicial polytope are given by the Dehn-Somerville relations, which takes a similar form to Pascal's triangle. For example, a 4-phase coexistence tetrahedron is a 3-dimensional simplex, which has $_4C_1 = 4$ vertices (single phases), $_4C_2 = 6$ edges (2-phase coexistence), $_4C_3 = 4$ triangles (3-phase) and $_4C_4 = 1$ tetrahedron (4-phase). Even when one performs a Legendre transformation, the fundamental underlying geometric structure of the $U(S,X)$ simplex, as well as its coexistence regions, are preserved.

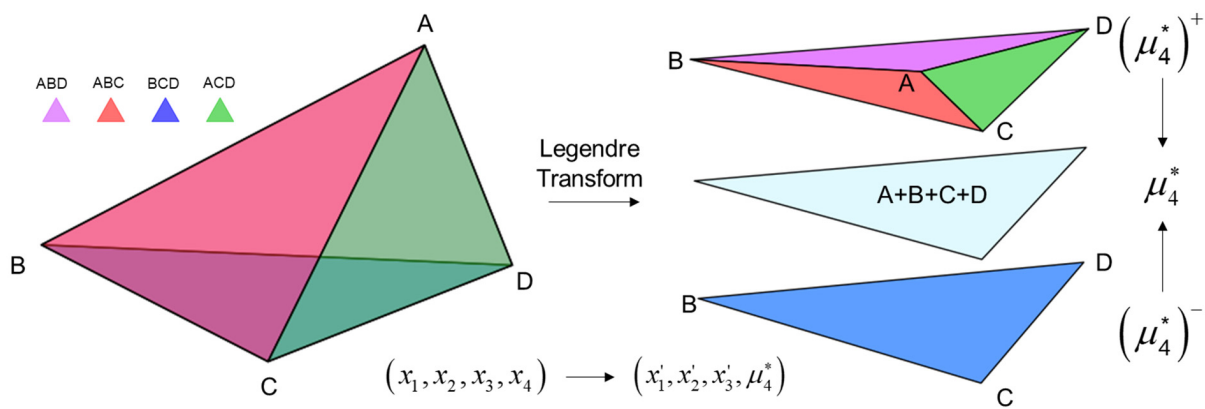

**Figure 6**. A 3D simplicial polytope (a tetrahedron) with extensive natural variables is smushed via a Legendre transformation to a fixed intensive variable of $\mu_4^*$. All simplicial facets from the tetrahedron, and thereby its phase coexistence information, remains preserved following the Legendre transformation.

On a four-phase coexistence tetrahedron, all vertices share the same chemical potentials $\mu_1$ through $\mu_4$. On composition axes, the coordinates for each vertex can be written as $(x_1, x_2, x_3, x_4)$, where 1, 2, 3, 4 correspond to different elements, and $x_4 = 1 - x_1 - x_2 - x_3$ by the affine constraint. Upon a Legendre transformation from $x_4$ to $\mu_4$, the $x_4$ coordinates all change to the same $\mu_4$, which effectively smushes all the vertices, edges and facets from the tetrahedron onto a single $\mu_4^*$ value, as illustrated in **Figure 6**. Each $x_i$ coordinate then changes to $x_i'$ by the new affine constraint, $x_3' = 1 - x_1' - x_2'$.

Importantly, all the 2-phase edges, 3-phase triangles, and the 4-phase tetrahedron are preserved after the Legendre transformation. The Legendre transformation does not generate any new phase coexistence information, nor does it lose any information. It simply provides a different perspective, but for the same equilibrium of heterogeneous substances. On mixed phase diagrams the phase coexistence regions no longer appear like simplicial polytopes, but are in fact down-projections of the high-dimensional simplices from the $U(S,X_i)$ space where they originated. In the **SI5**, we provide tables that explicitly describe the dimensionality of coexistence regions on mixed phase diagrams. These the geometric considerations of phase coexistence are the same with any other intensive variable as well, for example on a traditional ternary composition phase diagram $(x_1, x_2, x_3, T)$, or replacing $\mu_4$ with pressure, magnetic field, area-to-volume ratio, *etc.*

**Duality in Thermodynamics**

Duality gives *two different points of view of looking at the same object.* As summarized in the table below, there is a duality in thermodynamics between open and closed systems; which corresponds to a duality between the Internal Energy potential and its Legendre transformation; which corresponds to a duality in computation between a convex hull and its half-space intersection. Our implementation of these duality concepts for chemical work are geometrically identical to the duality relationships between the Maxwell $U(S,V)$ surface and the $G(T,P)$ free energy surfaces which we are commonly familiar with today. Our primary contribution here was to extend these concepts from the $G(T,P,x)$ space to the grand potential $\phi(T,P,\mu)$ space. The dualities in macroscopic boundary conditions for classical thermodynamics can further be linked to dualities at the atomistic scale, in the statistical mechanics description between an $NVT$ canonical ensemble and the $\mu VT$ grand canonical ensemble.

| **Duality in Thermodynamics** | | |
|---|---|---|
| Thermodynamic System | Closed equilibrium mixture of heterogeneous substances | Subsystem of single material open to an external reservoir |
| Thermodynamic Potential | Internal Energy $U(X_i)$ with extensive natural variables | Legendre transformation to $\Phi(Y_i)$ with natural intensive variable |
| Computational Thermodynamics | Convex hull of vertices | Half-space intersection of hyperplanes |

Our motivation for this work was to address the underutilization of chemical potential diagrams in the existing materials thermodynamics literature. Although computational tools for chemical potential diagrams have existed for over two decades, we believe that the bottleneck to their widespread proliferation is not their computation, but rather, is the physical understanding and interpretation of these diagrams. In particular, chemical potential diagrams offer a unique connection to the kinetics of diffusion, nucleation and growth, which has broad and obvious value in materials science and engineering. Therefore, the essential intellectual task in deploying chemical potential diagrams (or any phase diagram for that matter) is connecting a physical scenario to its boundary conditions and corresponding thermodynamic potential, and then from the available thermochemical data to the computation of a final phase diagram.

In framing open boundary conditions to analyze the stability of a material-of-interest with respect to an open reservoir, the next question becomes, how can one control the relative stability of a target material, when there may be numerous forms of available work? In Part III of this series, we derive a generalized Clausius-Clapeyron relation to examine the gradients of phase boundaries on high-dimensional phase diagrams, providing a pathway to control the relative stability of specific phases.

## Code Availability

All code for analyzing and visualizing convex hull, high dimensional chemical potential diagrams, and mixed composition-chemical potential diagrams can be found on Github at the following link:

https://github.com/dd-debug/chemical_potential_diagram_and_convex_hull_and_pourbaix_diagram

The link includes a readme, tutorial example files, installation guide, Python package requirements, and instructions for use.

## Acknowledgements

This work was supported as part of GENESIS: A Next Generation Synthesis Center, an Energy Frontier Research Center funded by the US Department of Energy, Office of Science, Basic Energy Sciences (award No. DE-SC0019212).